\documentclass[aps,pre,twocolumn]{revtex4}

\usepackage{amssymb}
\usepackage{graphicx}
\usepackage{color}
\usepackage{comment}

\begin{document}

\title{Temporal fluctuation scaling in nonstationary counting processes}
\author{Shinsuke Koyama} 
\email{skoyama@ism.ac.jp}
\affiliation{Department of Statistical Modeling, 
The Institute of Statistical Mathematics, 
10-3 Midori-cho, Tachikawa, Tokyo 190-8562, Japan}
\date{\today}

\begin{abstract}
The fluctuation scaling law has universally been observed in a wide variety of phenomena.
For counting processes describing the number of events occurred during time intervals, 
it is expressed as a power function relationship between the variance and the mean of the event count per unit time, the characteristic exponent of which is obtained theoretically in the limit of long duration of counting windows.
Here I show that the scaling law effectively appears even in a short timescale in which only a few events occur. 
Consequently, the counting statistics of nonstationary event sequences are shown to exhibit the scaling law as well as the dynamics at temporal resolution of this timescale. 
I also propose a method to extract in a systematic manner the characteristic scaling exponent from nonstationary data.
\end{abstract}

\maketitle


The fluctuation scaling law 
has been observed in many natural and man-made systems. 
It was originally found by Taylor in ecological systems as an empirical power function relationship between the variance and the mean of the number of individuals of a species per unit area \cite{Taylor61}. 
The scaling relationship has been demonstrated in other fields such as transmission of infectious diseases, cancer metastasis, chromosomal structure and traffic in transportation networks \cite{Anderson89,Kendal87,Kendal04,Fronczak10,Barabasi04}, showing a universality of the law.

This letter focuses particularly on 
the fluctuation scaling law in counting processes. 
A counting process is a stochastic process $\{N_t; t\ge1\}$ describing  
the number of events occurred in the interval $(0,t]$, 
which is used for modeling a wide variety of phenomena 
such as occurrence of earth quake, photon counting and neural spike trains \cite{Ogata88,Scully97,Johnson96}.  
The fluctuation scaling law for the counting process considered here states that the variance of $N_t$ per unit time is a power function of the mean of $N_t$ per unit time,
\begin{equation}
Var(N_t)/t \propto [E(N_t)/t]^{\beta}.
\label{eq:fs-counting}
\end{equation}
Since a random process leads to a Poisson process, $\beta=1$ becomes an indicator of randomness: every deviation from randomness indicates a deviation from this relationship. 

To compute the mean and the variance of $N_t$, 
it is usually taken a counting window of long duration $t\gg1$ in which a large number of events occur.
However, the scaling law (\ref{eq:fs-counting}) generally depends on the duration of the counting window, or on the average number of events in the window. 
In the limit of $t\to0$, for instance, it can be shown that the scaling relation with arbitrary exponent vanishes but $\beta$ approaches unity, 
which is essentially the same as the fact that the Fano factor $F(t):=Var(N_t)/E(N_t)$ for any regular point process approaches unity for $t\to0$ \cite{Teich97}.
It is therefore of interest how many events on average in the counting window is enough to observe the scaling law, 
the exponent of which characterizes the `intrinsic' variability of occurrence of events. 

This question is important particularly for nonstationary sequences of events. 
In nervous systems, for example, the firing rate is typically modulated with timescale of tens to hundreds milliseconds, in which only a few events (spikes) occur \cite{Richmond87}. 
Since neurons operate in such a short timescale, 
it is important to ask if the counting statistics exhibits the scaling law with only a few events. 

Here, I show by assuming renewal processes that the scaling law in the counting statistics appears even in a short counting window, in which only a few events on average occur. 
I also propose a method to extract in a systematic manner the characteristic scaling exponent 
from nonstationary sequences of events. 
The ability of the proposed method is demonstrated with data simulated by a leaky integrate-and-fire neuron model.


I begin with the fluctuation scaling law for stationary renewal processes.
Let $X\ge0$ be an interevent interval, and $m=E(X)$ and $s^2=Var(X)$ be its mean and variance, respectively. 
Suppose that the variance has a power function relation with $m$ as 
\begin{equation}
s^2 = \kappa m^{\alpha}.
\label{eq:scaling-isi}
\end{equation}
The scaling exponent $\alpha$ characterizes the `intrinsic' dispersion of occurrence of events. 
For a Poisson (random) process, $\alpha=2$. 
On the other hand, $\alpha>2$ $(<2)$ implies the tendency for the timing of event occurrence to be over (under) dispersed for large means, and under (over) dispersed for small means.

Let $N_t$ be the number of events occurred in the counting window $(0,t]$.
For $t\gg1$, $N_t$ asymptotically follows the Gaussian distribution with mean $t/m$ and variance $s^2t/m^3$ \cite{Cox62}.
Then, if the interval statistics has the scaling property (\ref{eq:scaling-isi}), 
the variance of $N_t$ per unit time is asymptotically scaled by the mean of $N_t$ per unit time (i.e., the rate) as 
\begin{equation}
\frac{Var(N_t)}{t} = \kappa\bigg[\frac{E(N_t)}{t}\bigg]^{\beta},
\label{eq:scaling-count}
\end{equation}
where $\beta=3-\alpha$ for $t\gg 1$.
Note that the scaling law (\ref{eq:scaling-count}) depends on the duration $t$ of the counting window. 
In theory, the exponent $\beta=3-\alpha$ is achieved in the limit of $t\to\infty$, in which a sufficiently large number of events occur.
On the other hand, $\beta$ approaches 1 for the limit of $t\to0$.

One can construct an interevent interval density of renewal process that possesses the scaling law (\ref{eq:scaling-isi}) by introducing a parametric probability density $f(z;\kappa)$ with unit mean and the variance $\kappa$, and rescaling it as 
\begin{equation}
p(x;\lambda,\kappa,\alpha) = \lambda f(\lambda x;\lambda^{2-\alpha}\kappa),
\label{eq:scaling-density}
\end{equation}
where $\lambda=1/m$ is the frequency of occurrence of events.
Here, the choice of $f(z;\kappa)$ is arbitrary: 
different choices of $f(z;\kappa)$ generate different families of probability densities that have the scaling law (\ref{eq:scaling-isi}).

A nonstationary renewal process that exhibits the scaling law in the counting statistics can be constructed by generalizing the construction of nonstationary Poisson processes, 
the idea of which is given as follows. 
Consider a stationary Poisson process with unit rate defined on dimensionless time $s$. 
Then, the probability of occurring an event in a short interval $(s,s+ds]$ is given by $ds$.
Let $\lambda(t)$ be a rate of event occurrence on the real time $t$ and $\Lambda(t)=\int_0^t\lambda(u)du$ be the cumulative function of $\lambda(t)$.
By transforming the time $s$ into $t$ with $s = \Lambda(t)$, one obtain a nonstationary Poisson process with time-dependent rate $\lambda(t)$, in which the probability of occurring an event in a short interval $(t,t+dt]$ is $\lambda(t)dt$.
In the same manner, any renewal process with unit rate can be transformed by $s=\Lambda(t)$ into a nonstationary renewal process with the trial-averaged rate $\lambda(t)$ \cite{Berman81,Barbieri01,Koyama08, Pillow08}. 
However, this transformation does not allow the variance of the event count per unit time to have the power function of the rate with arbitrary scaling exponent \footnote{
In fact, this transformation results in the variance of the event count being proportional to the mean, $Var(N_t) = \kappa E(N_t)$ for $t\gg1$.
}.

Hence, I propose a generalization of the transformation so that the variance and the mean of count per unit time obey the scaling law (\ref{eq:scaling-count}).
Consider a renewal process with the interevent interval density $f(z;\kappa)$. 
The conditional rate, or the hazard funciton, of this process is given by 
\begin{equation}
g(s;s_*,\kappa) = \frac{f(s_i-s_*;\kappa)}{1-\int_{s_*}^sf(u-s_*;\kappa)du},
\label{eq:hazard}
\end{equation}
where $s_*(<s)$ is the last event time preceding $s$.
Analogously to Eq.~(\ref{eq:scaling-density}), 
by rescaling the variance parameter $\kappa\to\lambda(t)^{2-\alpha}\kappa$ as well as the time $s = \Lambda(t)$, the conditional rate of the nonstationary renewal process is obtained as 
\begin{eqnarray}
\lefteqn{r(t;t_*,\{\lambda(t)\},\kappa,\alpha)}\hspace{0.5cm}\nonumber\\
&=&
\frac{\lambda(t)f(\Lambda(t)-\Lambda(t_*);\lambda(t)^{2-\alpha}\kappa) }
{1-\int_{t_*}^t\lambda(v)f(\Lambda(v)-\Lambda(t_*);\lambda(v)^{2-\alpha}\kappa)dv} .
\label{eq:cif}
\end{eqnarray}
For $dt\ll 1$, Eq.~(\ref{eq:cif}) gives the conditional probability of occurring an event in $(t,t+dt]$, given the last event at $t_*$, 
\begin{eqnarray}
\lefteqn{P(N_{t+dt}-N_t=1;t_*,\{\lambda(t)\},\kappa,\alpha)}\hspace{1cm}\nonumber\\
 &\approx&
  r(t;t_*,\{\lambda(t)\},\kappa,\alpha)dt,
\label{eq:prob-event}
\end{eqnarray}
which can be used for simulating sequences of events.


With the basis of the model (\ref{eq:cif}), I propose a method for estimating the scaling exponent $\alpha$ (and the coefficient $\kappa$) from data consisting of nonstationary sequences of events. 
The likelihood function of $(\alpha,\kappa)$, given a sequence of event times $\{t_i\}:=\{t_1,\ldots,t_n\}$, 
is expressed with the conditional rate function (\ref{eq:cif}) as  
\begin{eqnarray}
\lefteqn{l(\kappa,\alpha;\{t_i\},\{\lambda(t)\})}\hspace{0.1cm}\nonumber\\
 &=& 
\Bigg[\prod_{i=2}^n r(t_i;t_{i-1},\{\lambda(t)\},\kappa,\alpha) \Bigg] \nonumber\\
& & { } \times
\exp\Bigg(
-\int_{t_1}^{t_n} r(u;t_{N_u},\{\lambda(t)\},\kappa,\alpha) du
\Bigg) ,
\label{eq:likelihood1}
\end{eqnarray}
where the exponential factor represents the probability of no event in each interevent interval \cite{Daley02,Kass01}.
Substituting Eq.~(\ref{eq:cif}) into Eq.~(\ref{eq:likelihood1}), it can be expressed in more tractable form,
\begin{eqnarray}
\lefteqn{l(\kappa,\alpha;\{t_i\},\{\lambda(t)\})}\hspace{1cm}\nonumber\\ 
&=&
\prod_{i=2}^n \lambda(t_i) f(\Lambda(t_i)-\Lambda(t_{i-1});\lambda(t_i)^{2-\alpha}\kappa) .
\label{eq:likelihood2}
\end{eqnarray}
For $M$ independent and identically distributed trials $\{t_i^j\}_{j=1}^M:=\{t_1^j,\ldots,t_{n_j}^j\}_{j=1}^M$, $n_j$ denoting the number of events in the $j$th trial,
the likelihood function is simply given by the product of the likelihood function of single trials (\ref{eq:likelihood1}). 
Using this, the parameters can be estimated in the following two steps. 
1) Compute an estimate $\hat{\lambda}(t)$ of the trial-averaged rate function from $\{t_i^j\}_{j=1}^M$, which can be obtained by a kernel density estimator with a Gaussian kernel whose band-width is determined by minimizing the expected mean squared error \cite{Shimazaki10};
2) Substitute $\hat{\lambda}(t)$ and $\{t_i^j\}_{j=1}^M$ into the likelihood function, and maximize it with respect to $(\alpha,\kappa)$ to obtain the estimate $(\hat{\alpha},\hat{\kappa})$.

\begin{figure}[tb]
\begin{center}
\includegraphics[width=8.5cm]{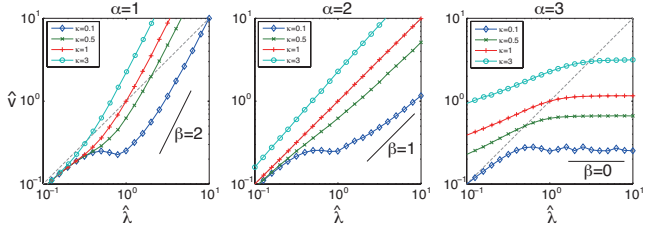}
\caption{
The log-log plot of the variance of event count $\hat{v}$ as a function of the mean $\hat{\lambda}$ for $\alpha=1$(left), 2(middle) and 3(right). The dashed lines of unit slope (indicating $\hat{v}=\hat{\lambda}$) are included for comparison.
Even with a few events on average, $\hat{v}$ and $\hat{\lambda}$ exhibit the scaling law with the exponent $\beta=3-\alpha$.
}
\label{fig:stationary}
\end{center}
\end{figure}

To study the behavior of the statistical model (\ref{eq:cif}), 
the stationary case (i.e., $\lambda(t)=\lambda$ is constant in time) is examined firstly.
For this purpose, the gamma density, 
\begin{equation}
f(z;\kappa) = \kappa^{-1/\kappa}z^{1/\kappa-1}e^{-z/\kappa}/\Gamma(1/\kappa),
\label{eq:gamma}
\end{equation}
is employed for $f(z;\kappa)$, and $M=10^5$ sequences of events 
are simulated using the interivent interval density (\ref{eq:scaling-density}), or equivalently the conditional rate function (\ref{eq:prob-event}) with $\lambda(t)=\lambda$, under the equilibrium condition for each $\alpha=1$, 2 and 3.
The equilibrium condition is ensured by starting the simulations some times before the actual measurement begins. 
The mean $\hat{\lambda}$ and the variance $\hat{v}$ of the number of event in the counting window of duration $\Delta=1$ are calculated from the $M$ trials, and are plotted on a log-log scale (Figure \ref{fig:stationary}). 
It is seen from this figure that 
$\hat{v}$ and $\hat{\lambda}$ asymptotically obey the scaling law $\hat{v}\propto\hat{\lambda}^{\beta}$ with $\beta=3-\alpha$ as $\hat{\lambda}$ is increased.
Recall that the scaling law (\ref{eq:scaling-count}) for stationary renewal processes is theoretically derived for a long duration of the counting window in which a large number of events occur, so that the central limit theorem can be applied \cite{Cox62}. 
The simulation result, however, shows that the scaling law with $\beta=3-\alpha$ appears even with a few events.

\begin{figure}[tb]
\begin{center}
\includegraphics[width=8.5cm]{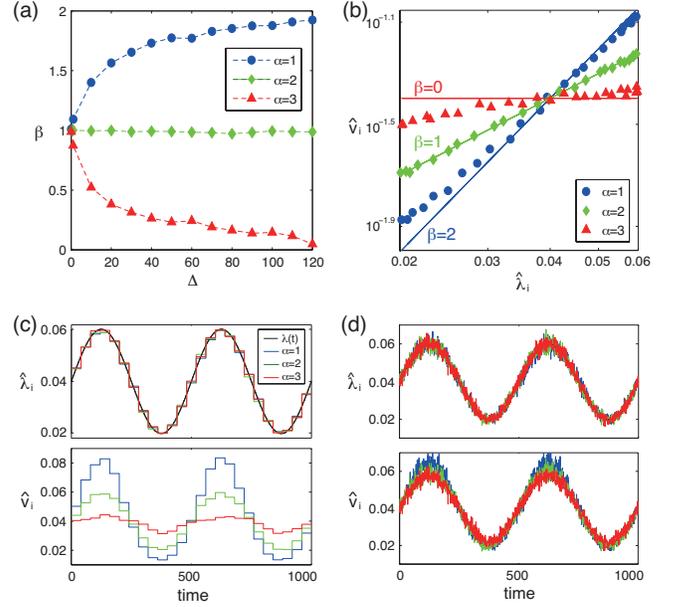}
\caption{
The results of the nonstationary case. The simulations are performed with the parameter values $\alpha=$1, 2 and 3, and $\kappa=0.04^{\alpha-2}$. 
In all the figures, results for $\alpha=$1, 2 and 3 are represented with blue, green and red, respectively.
(a) The slope $\beta$, obtained by the linear regression of $\{\log\hat{v}_i\}$ on $\{\log\hat{\lambda}_i\}$, as a function of $\Delta$.
(b) The log-log plot of $\hat{v}_i$ against $\hat{\lambda}_i$ calculated with $\Delta=40$.
The solid lines represent the theoretical slope $\beta=3-\alpha$.
(c) The time course of $\hat{\lambda}_i$ (top) and $\hat{v}_i$ (bottom) calculated with $\Delta=40$.
(d) The same as in (c) but with $\Delta=1$.
}
\label{fig:nonstationary}
\end{center}
\end{figure}

To examine the nonstationary case, a periodic function $\lambda(t) = 0.04 + 0.02\sin\frac{2\pi}{500}t$
is used  for the time-dependent rate, and $M=10^4$ sequences of events are simulated using Eq.~(\ref{eq:prob-event}) in the time interval $t\in(0,1000]$ for each $\alpha$=1, 2 and 3.
The time axis is divided into equally spaced, contiguous time windows, each of duration $\Delta$, and the number of events in the $i$th window of the $j$th trial is counted and denoted by $N_i^j$.
The mean and the variance of the event count per unit time in the $i$th window are respectively computed as
\begin{equation}
\hat{\lambda}_i=\frac{ \frac{1}{M}\sum_{j=1}^MN_i^j }{\Delta},
\label{eq:count-mean}
\end{equation}
and
\begin{equation}
\hat{v}_i = \frac{ \frac{1}{M-1}\sum_{j=1}^M(N_i^j - \hat{\lambda}_i\Delta)^2 }{\Delta}.
\label{eq:count-variance}
\end{equation}
The slope $\beta$ is then computed by performing the linear regression of $\{\log \hat{v}_i\}$ on $\{\log\hat{\lambda}_i\}$.
Figure~\ref{fig:nonstationary}a depicts $\beta$ as a function of $\Delta$, showing that 
$\beta$ approaches $3-\alpha$ as $\Delta$ is increased while $\beta\to1$ as $\Delta\to0$.
For illustration, Figure~\ref{fig:nonstationary}b plots $\{\hat{v}_i\}$ against $\{\hat{\lambda}_i\}$, which are calculated with $\Delta=40$, on a log-log scale, in which 
it is seen that the scaling law with the exponent $\beta=3-\alpha$ approximately holds for 
relatively large $\hat{\lambda}_i$ (i.e., the average number of events in the counting windows is roughly more than 1), which correlates with the finding in the stationary case. 
With the time resolution $\Delta=40$, $\hat{v}_i$s are dynamically modulated in proportion to $\hat{\lambda}_i^{\beta}$s with $\beta\approx3-\alpha$ (Figure~\ref{fig:nonstationary}c).
If the duration of counting windows is taken to be $\Delta=1$, the event count exhibits nearly the Poisson variance (i.e., $\hat{v}_i=\hat{\lambda}_i$), so that $\hat{v}_i$s of the three cases are hardly distinguishable from each other (Figure~\ref{fig:nonstationary}d).

The ability of the proposed inference method to extract the characteristic scaling exponent $\alpha$ is tested with data simulated by a leaky integrate-and-fire (LIF) neuron model. 
For this purpose, I first examine the scaling property of spike trains generated from the LIF model.
The dynamics of the LIF model are represented by the equation \cite{Koch99},
\begin{equation}
\frac{dV(t)}{dt} = -\frac{V(t)}{\tau} + I(t),
\end{equation}
where $V(t)$ is the membrane potential, $\tau=5$ is the membrane decay time constant, and $I(t)$ represents the input current. When the membrane potential reaches the threshold $v_{th}=1$, an event (spike) is generated and the membrane potential is reset to 0 immediately.

\begin{figure}[tb]
\begin{center}
\includegraphics[width=8.5cm]{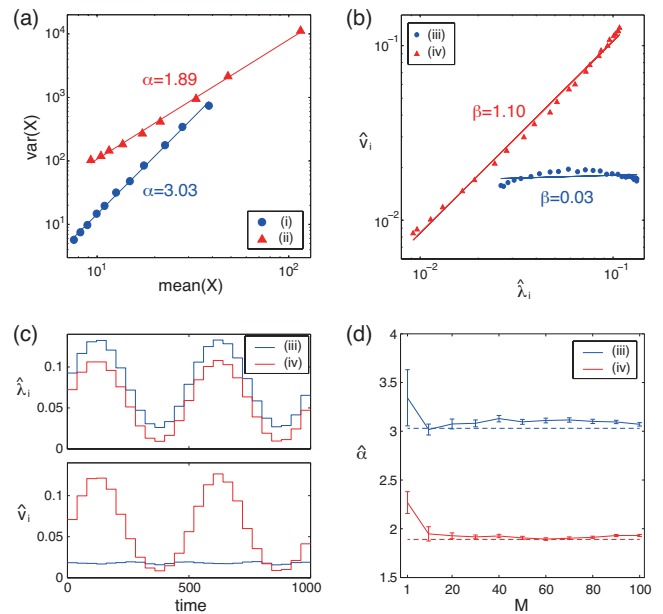}
\caption{
The result of the LIF model.
In all the figures, blue indicates that the mean current $\mu$ varies, while red indicates that the amplitude of current fluctuation $\sigma$ varies.
(a) The log-log plot of the variance against the mean of interevent interval when the stationary input currents are injected. The solid lines represent the slopes obtained by the linear regression on the log-log scale. 
(b) The log-log plot of $\hat{v}_i$ against $\hat{\lambda}_i$ calculated with $\Delta=40$ when the nonstationary current inputs are injected. 
The solid lines represent the slopes obtained by the linear regression on the log-log scale. 
(c) The time courses of $\hat{\lambda}_i$ (top) and $\hat{v}_i$ (bottom) calculated with $\Delta=40$.
(d) The estimate $\hat{\alpha}$ as a function of the number of trials $M$. 
The standard deviations are computed with 100 realizations.
}
\label{fig:lif}
\end{center}
\end{figure}

For a stationary input current $I(t) = \mu + \sigma \xi(t)$, 
where $\xi(t)$ is a Gaussian white noise satisfying $\langle\xi(t)\rangle=0$ and $\langle\xi(t)\xi(t')\rangle=\delta(t-t')$, 
the following two cases are examined in the simulations:
(i) $\mu$ varies from 0.15 to 0.25 
while $\sigma=0.1$ is fixed, and (ii) $\sigma$ varies from 0.2 to 0.6 
while $\mu=0.05$ is fixed. 
For each set of parameter values $(\mu,\sigma)$, a sequence of $10^3$ events is generated, from which the mean and the variance of interevent interval are calculated. 
It is found that the variance and the mean obey the scaling law,
whose exponent obtained by the linear regression on a log-log scale is 
$\alpha=3.03$ for (i) and $\alpha=1.89$ for (ii) (Figure~\ref{fig:lif}a).

For a nonstationary input current $I(t) = \mu(t) + \sigma(t)\xi(t)$, the following two cases are considered analogously to the stationary case:
(iii) the mean current has a periodic profile $\mu(t)=0.2+0.05\sin\frac{2\pi }{500}t$, while the amplitude of the current fluctuation is constant $\sigma(t)=0.1$, and 
(iv) the amplitude of the current fluctuation is periodically modulated $\sigma(t)=0.4+0.2\sin\frac{2\pi }{500}t$, while the mean current is constant $\mu(t)=0.05$.
For each case, $M=10^4$ spike trains are simulated in the time interval $t\in(0,1000]$, from which $\{\hat{\lambda}_i\}$ and $\{\hat{v}_i\}$ are computed by Eqs.~(\ref{eq:count-mean})-(\ref{eq:count-variance}) with $\Delta=40$. 
Figures~\ref{fig:lif}b and c depict the result, showing that $\hat{v}_i$ and $\hat{\lambda}_i$ approximately obey the scaling law.
The slope obtained by the linear regression of $\{\log\hat{v}_i\}$ on $\{\log\hat{\lambda}_i\}$ is $\beta=0.03$ for (iii) and $\beta=1.10$ for (iv), from which we see the approximate relation $\beta\approx3-\alpha$.
It is, thus, empirically confirmed that the spike trains generated from the LIF model exhibit the scaling law, which qualitatively matches that of the statistical model (\ref{eq:cif}) (Figures~\ref{fig:nonstationary}b and c).

I finally examine if the proposed inference method can capture the scaling exponent $\alpha$ directly from the nonstationary sequences of events generated by the LIF model.
For each nonstationary input current (iii) and (iv), 
the inference method is applied to $M$ spike trains simulated by the LIF model to obtain $\hat{\alpha}$.
Figure~\ref{fig:lif}d plots $\hat{\alpha}$ against the number of trials $M$, which shows that 
the accuracy of the estimation is improved as $M$ is increased. 
For example, the exponent is estimated from $M=20$ trials as 
$\hat{\alpha}= 3.08 \pm 0.05$ for the case (iii) and $\hat{\alpha}= 1.93 \pm 0.03$ for the case (iv), 
which are in good agreement with the values obtained in Figure~\ref{fig:lif}a.

For summary, it was shown in this letter that assuming renewal processes, only a few events in counting windows are enough for the variance and the mean per unit time to exhibit the scaling law with the exponent $\beta=3-\alpha$.
As a result, the counting statistics of nonstationary event sequences display the scaling law as well as the dynamics at temporal resolution of this counting windows (Figures~\ref{fig:nonstationary}c and \ref{fig:lif}c). 
I also proposed a method based on the likelihood principle to extract the scaling exponent from nonstationary sequences of events, the ability of which was demonstrated with the data simulated by the LIF model. 

The results of renewal processes can be generalized to nonrenewal processes directly.  
For nonrenewal processes, whose interval statistics has the scaling law (\ref{eq:scaling-isi}), the asymptotic scaling relation in the counting statistics (\ref{eq:scaling-count}) remains unchanged except that the coefficient is modified \footnote{
For stationary point processes with serially correlated interevent intervals and the scaling relation (\ref{eq:scaling-isi}), it holds for $t\gg1$ that  
$Var(N_t)/t = \kappa(1+2\eta)[E(N_t)/t]^{3-\alpha}$ with $\eta=\sum_{i=1}^{\infty}\eta_i$, where $\eta_i$ denotes the $i$th-order linear correlation coefficient for pairs of intervals that are separated by $i-1$ intermediate intervals \cite{Nawrot10}.
}. 
The generalization of the proposed inference method to the nonrenewal processes is also straightforward since the transformation of a stationary point process (\ref{eq:hazard}) to a nonstationary point process (\ref{eq:cif}) is applicable to nonrenewal processes.

In nervous systems, neurons produce an action potential by integrating presynaptic inputs within tens milliseconds, in which typically only a few spikes come from each presynaptic neuron. 
This suggests that the scaling law in spike count effectively appears in the integration time, and thus may have an impact on information processing.
Ma {\it et al.} \cite{Ma06} suggested a hypothesis that the Poisson-like statistics in the responses of populations of cortical neurons may represent probability distributions over the stimulus and implement Bayesian inferences. 
An important property in their hypothesis is that the variance of spike count is proportional to the mean spike count, which corresponds to $\beta=1$ (or $\alpha=2$ in the interval statistics) in our formulation.
It is worth pointing out that $\beta\approx1$ is observed in the simulations of the LIF neuron with the fluctuating current input (Figure~\ref{fig:lif} red), which can be realized by balanced excitatory and inhibitory synaptic inputs observed in the cortex \cite{Vreeswijk96,Amit97,Destexhe01,Shu03,Haider06}. 

On the other hand, from {\it in vivo} recordings, Troy and Robson \cite{Troy92} found that steady discharges of retinal ganglion cells, in response to stationary visual patterns, exhibit the scaling law in the interval statistics (\ref{eq:scaling-isi}) with exponent $\alpha\approx 3$. 
This exponent is also observed in the simulations using multi-compartment models of retinal ganglion cells \cite{Rossum03} as well as the LIF model with the current input whose mean is modulated (Figure~\ref{fig:lif}a blue).

It is therefore speculated that the scaling exponent $\alpha$ may reflect the intrinsic mechanisms of neuronal discharge or the internal dynamics of networks \cite{Barabasi04-2}, and may be related to schemes for neural computation the nervous systems employ. 
The proposed method offers a systematic way to extract the characteristic scaling exponent from experimental data.

This research was supported by JSPS KAKENHI Grant Number 24700287.


\end{document}